\documentclass[journal=jacsat,manuscript=article]{achemso}

\usepackage{chemformula} % Formula subscripts using \ch{}
\usepackage{bm, color}
\usepackage{bbm}
\usepackage{lineno}
\usepackage{braket}
\usepackage{amsbsy}
\usepackage{xr}

\author{Junwen Lai}
\affiliation{School of Materials Science and Engineering, University of Science and Technology of China, Shenyang 110016, China.}
\alsoaffiliation{Shenyang National Laboratory for Materials Science, Institute of Metal Research,Chinese Academy of Sciences, Shenyang 110016, China.}

\author{Xiangyang Liu}
\affiliation{School of Materials Science and Engineering, University of Science and Technology of China, Shenyang 110016, China.}
\alsoaffiliation{Shenyang National Laboratory for Materials Science, Institute of Metal Research,Chinese Academy of Sciences, Shenyang 110016, China.}

\author{Jie Zhan}
\affiliation{School of Materials Science and Engineering, University of Science and Technology of China, Shenyang 110016, China.}
\alsoaffiliation{Shenyang National Laboratory for Materials Science, Institute of Metal Research,Chinese Academy of Sciences, Shenyang 110016, China.}

\author{Tianye Yu}
\affiliation{Shenyang National Laboratory for Materials Science, Institute of Metal Research,Chinese Academy of Sciences, Shenyang 110016, China.}

\author{Peitao Liu}
\affiliation{School of Materials Science and Engineering, University of Science and Technology of China, Shenyang 110016, China.}
\alsoaffiliation{Shenyang National Laboratory for Materials Science, Institute of Metal Research,Chinese Academy of Sciences, Shenyang 110016, China.}

\author{Xing-Qiu Chen}
\email{xingqiu.chen@imr.ac.cn}
\affiliation{School of Materials Science and Engineering, University of Science and Technology of China, Shenyang 110016, China.}
\alsoaffiliation{Shenyang National Laboratory for Materials Science, Institute of Metal Research,Chinese Academy of Sciences, Shenyang 110016, China.}

\author{Yan Sun}
\email{sunyan@imr.ac.cn}
\affiliation{School of Materials Science and Engineering, University of Science and Technology of China, Shenyang 110016, China.}
\alsoaffiliation{Shenyang National Laboratory for Materials Science, Institute of Metal Research,Chinese Academy of Sciences, Shenyang 110016, China.}

\title{Switchable quantized signal between longitudinal conductance and Hall conductance in dual quantum spin Hall insulator TaIrTe$_4$}

\begin{document}
	
	%\begin{tocentry}
	
	%\end{tocentry}
	
	\begin{abstract}
		Topological insulating states in two-dimensional (2D) materials are ideal 
		systems to study different types of quantized response signals due to 
		their in gap metallic states. Very recently, the quantum spin Hall (QSH) effect
		was discovered in monolayer $\text{TaIrTe}_4$ via the observation of quantized
		longitudinal conductance that rarely exists in other 2D topological insulators.
		The non-trivial $Z_2$ topological charges can exist at both charge neutrality point
		and the van Hove singularity point with correlation effect induced band gap.  
		Based on this model 2D material, we studied the switch of quantized signals between
		longitudinal conductance and transversal Hall conductance via tuning external magnetic
		field. In $Z_2$ topological phase of monolayer $\text{TaIrTe}_4$, the zero Chern 
		number can be understood as 1-1=0 from the double band inversion from spin-up
		and spin-down channels. After applying a magnetic field perpendicular to the plane,
		the Zeeman split changes the band order for one branch of the band inversion from 
		spin-up and spin-down channels, along with a sign charge of the Berry phase.
		Then the net Chern number of 1-1=0 is tuned to 
		1+1=2 or -1-1=-2 depending on the orientation of the magnetic field. The quantized signal
		not only provides another effective method for the verification of topological state in 
		monolayer $\text{TaIrTe}_4$, but also offers a strategy for the utilization of 
		the new quantum topological states based on switchable quantized responses.
	\end{abstract}
	
	\section{Introduction and motivation}
	Topological states in two-dimensional (2D) systems have been extensively studied
	in the last decades. The quantum Hall (QH) effect is the first discovered topological
	state in materials, which presents as quantized Hall conductance in the unit
	of $e^{2}/h$ with zero longitudinal resistance\cite{PhysRevB.23.5632,PhysRevLett.48.1559,thouless1982,tsukazaki2007}. 
	The quantized Hall conductance
	originated from the dissipative chiral edge state, while all the other
	states are localized. In electronic band structures, the occupied and non-occupied states are connected by the chiral edge states located in the bulk band gap.
	The number of net edge channels can be understood from the Chern number of
	bulk band structures. Since the Hall conductance in the QH effect is only dependent on
	the fundamental constant of the electron charge and the Planck constant, 
	it plays an important role in the metrology resistance standards
	and quantum computing\cite{hasan2010, xiao2010,qi2011,RevModPhys.87.1213}.
	
	In history, most of the QH effect states were measured in
	2D electron gas under strong perpendicular magnetic fields
	\cite{PhysRevLett.63.1984,chu1992,jiang1993,liu2008,maciejko2011,feng2015}. 
	With the generalization of topological band theory in condensed matter physics,
	it is found that different types of topological states exist in nature
	and applied magnetic fields can control different topological phase transitions,
	including both insulating and semimetal states.
	With this guiding principle, the QH effect and quantum anomalous Hall (QAH) effect
	were realized in topological insulators and the thin film of Dirac 
	semimetals, \cite{yu2010,apalkov2011,zyuzin2011,beugeling2012,liu2013,zhang2013,bahari2016,chang2016,uchida2017,asaba2018,zhang2019,satake2020,kubisa2021,guo2023}
	and the QH effect is even generalized into three-dimensional (3D) electron 
	systems\cite{tang2019,galeski2020,PhysRevLett.125.036602,li2021,liu2021,li2021a}.
	
	The interplay between the magnetic field and 2D topological materials provides
	an ideal platform for the study of topological phase transition among 
	quantum spin Hall (QSH) insulators,
	topological semimetal, QH effect, and QAH effect et al\cite{kawamura2018,zhan2022,kong2022}.
	In addition to plenty of quantum topological phases, the topological phase transition
	also offers an effective approach for detecting the topological states from quantized
	transport signals. Very recently, TaIrTe$_4$ monolayer was experimentally fabricated
	and confirmed as a new dual QSH insulator\cite{tang2024}. 
	The nontrivial Z$_2$ topological charges in
	TaIrTe$_4$ exist at both the charge neutrality point and the van Hove singularity point
	with a new band gap induced by strong correlations.
	Owing to its 2D nature, monolayer TaIrTe$_4$ can be understood as a model material for
	realizing different types of topological states under perturbations of strain, gating, 
	magnetic field, and magnetic doping et al.
	
\begin{figure*}[htb]
		\centering
		\includegraphics[scale=0.15]{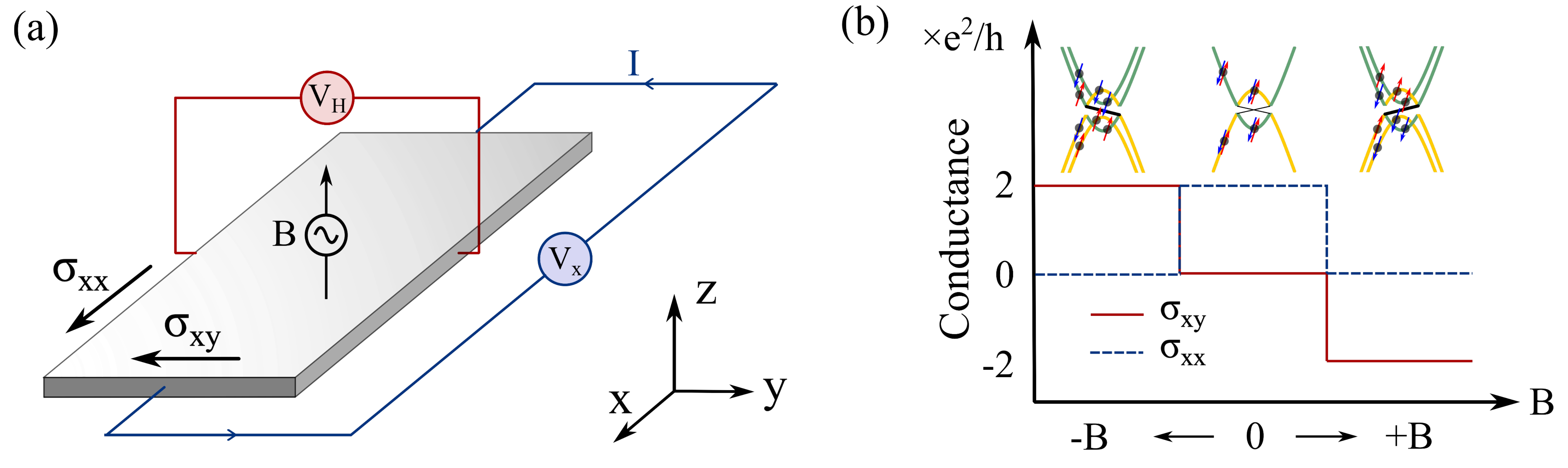}
		\caption{Schematic illustration of the magnetic field-induced conductance change 
			under topological phase transition.
			(a) Illustration of the longitude($\sigma_{xx}$) and Hall($\sigma_{xy}$) conductance 
			under a varying perpendicular magnetic field B, the input current I
			is along the $x$ direction.
			(b) Topological phase transition induced conductance
			($\sigma_{xx},\sigma_{xy}$) change under a varying perpendicular magnetic field B,
			the yellow and green colors stand for the different orbitals, the red and blue arrows
			on the band structure represents the spin up and down states, respectively.}
		\label{fig1}
	\end{figure*}

	In this work, we studied the evolution of magnetic field-induced topological phase 
	transition from Z$_2$ QSH insulator to QH insulator. Along with these phase transitions, 
	the quantized signals of longitudinal conductance and Hall conductance are switched on and 
	off via the control of an external magnetic field, as schematic shown in Fig.~\ref{fig1}. 
	The topological phase transition provides an effective approach to obtain a QH effect state 
	starting from the time-reversal symmetry 2D topological insulators, and
	the switchable quantized signals offer a strategy for the
	utilization of the new quantum topological material.
	
	\section{Result and Discussion}
\begin{figure}[htb]
		\centering
		\includegraphics[scale=0.12]{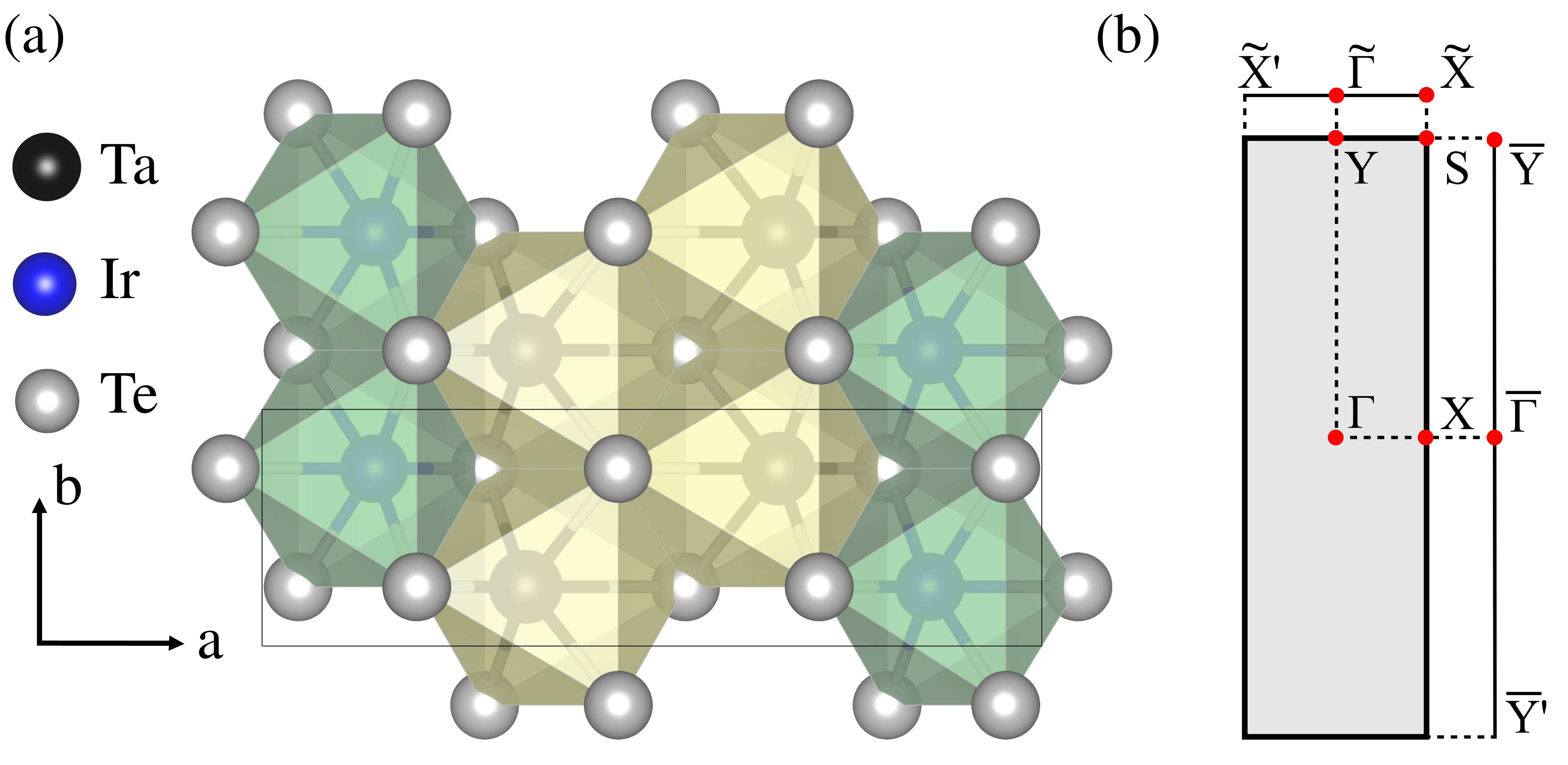}
		\caption{Crystal lattice structure and Brillouin zone of monolayer $\text{TaIrTe}_4$.
			(a) Top view of monolayer $\text{TaIrTe}_4$ 
			(b) Brillouin zone and its projection to different 
			direction [100] and [010] of monolayer $\text{TaIrTe}_4$}
		\label{fig2}
	\end{figure}

	Bulk $\text{TaIrTe}_4$ is a well known typical type-II Weyl semimetal 
	which crystallized by an AB stacking of two centrosymmetric van-der waals  
	layers\cite{mar1992,kumar2021}. Its monolayer was theoretically predicted as a QSH insulator
	at the charge neutrality point and very recently experimentally 
	verified by the observation of quantized longitudinal conductance~\cite{tang2024,guo2020}.
	In addition, the correlation effect induced a nontrivial $Z_2$ band gap
	also existed when weakly doping shifted the Fermi level to the van Hove 
	singularity points~\cite{tang2024}.The monolayer $\text{TaIrTe}_4$ crystallized in a 
	space group of $P2_1/m$ (No.11) which consists of two symmetry operators of inversion
	$i$ and \{$C_{2y}|(0,1/2,0)$\}, see Fig.~\ref{fig2}.
	
\begin{figure}[htb]
		\centering
		\includegraphics[scale=0.15]{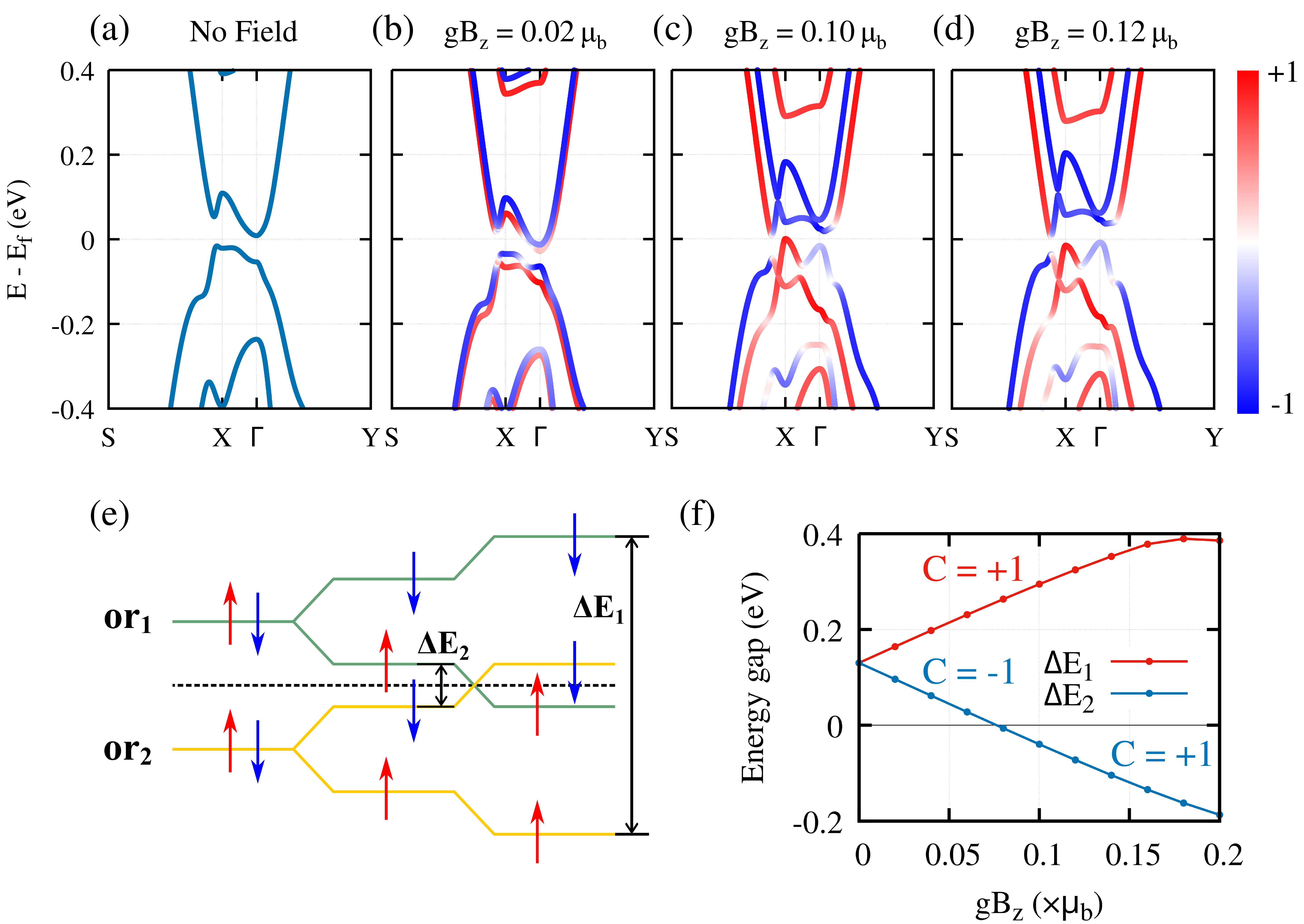}
		\caption{Evolution of electronic band structure and topological numbers of monolayer $\text{TaIrTe}_4$
			under magnetic field.
			(a) The band structure without magnetic field.
			(b-d) The spin-resolved band structures under different $gB_{z}$.
			The red and blue colors represent the z component of spin channels.
			(e) Schematic diagram of the band inversion progress at $X$ point with an increasing Zeeman field,
			the green and yellow lines stand for the different orbitals near Fermi energy
			,namely or1, and or2 ,
			the red and blue arrows stand for the spin states. 
			$\Delta E_1=E_{or1-down}-E_{or2-up}$ and $\Delta E_2=E_{or1-up}-E_{or2-down}$ are 
			energy differences between the spin up(down) state of or1 and spin down(up) state of or2.
			(f) The energy change of $\Delta E_1(\Delta E_2)$ at high symmetry point of $X$
			under an increasing magnetic field $gB_{z}$. $\Delta E_1$ states at positive zone in 
			the whole progress, with a constant Chern number 1. $\Delta E_2$ changes sign at around
			$gB_{z}=0.08~\mu_B$, along with the change of Chern number from -1 to 1.
		}
		\label{fig3}
	\end{figure}

	Based on the experimental reported lattice structure\cite{mar1992} , 
	the electronic band structures evolution of $\text{TaIrTe}_4$ is calculated
	with a tuning external magnetic field. 
	As presented in Fig.~\ref{fig3}(a), in the condition without external field, 
	there is a band inversion at $X$ point between the conduction and valence band with a 
	band gap of $\sim$24~meV,
	which can be seen from the ``W-shape" of the dispersion near the bottom of conduction bands,
	in good agreement with previous reports\cite{guo2020,zhao2022}.
	The $Z_2$ topological feature can be directly confirmed by the Wannier center evolution.  
	From Fig.~\ref{fig3}(a), one can easily see that the evolution of Wannier centers in 
	$k_{1}-k_{2}$ plane presents as a zigzag form with changing partners
	at the time-reversal invariant point. Hence, the evolution lines cross the reference line 
	an odd number of times in the half Brillouin zone (BZ).
	
\begin{figure}[htb]
		\centering
		\includegraphics[scale=0.15]{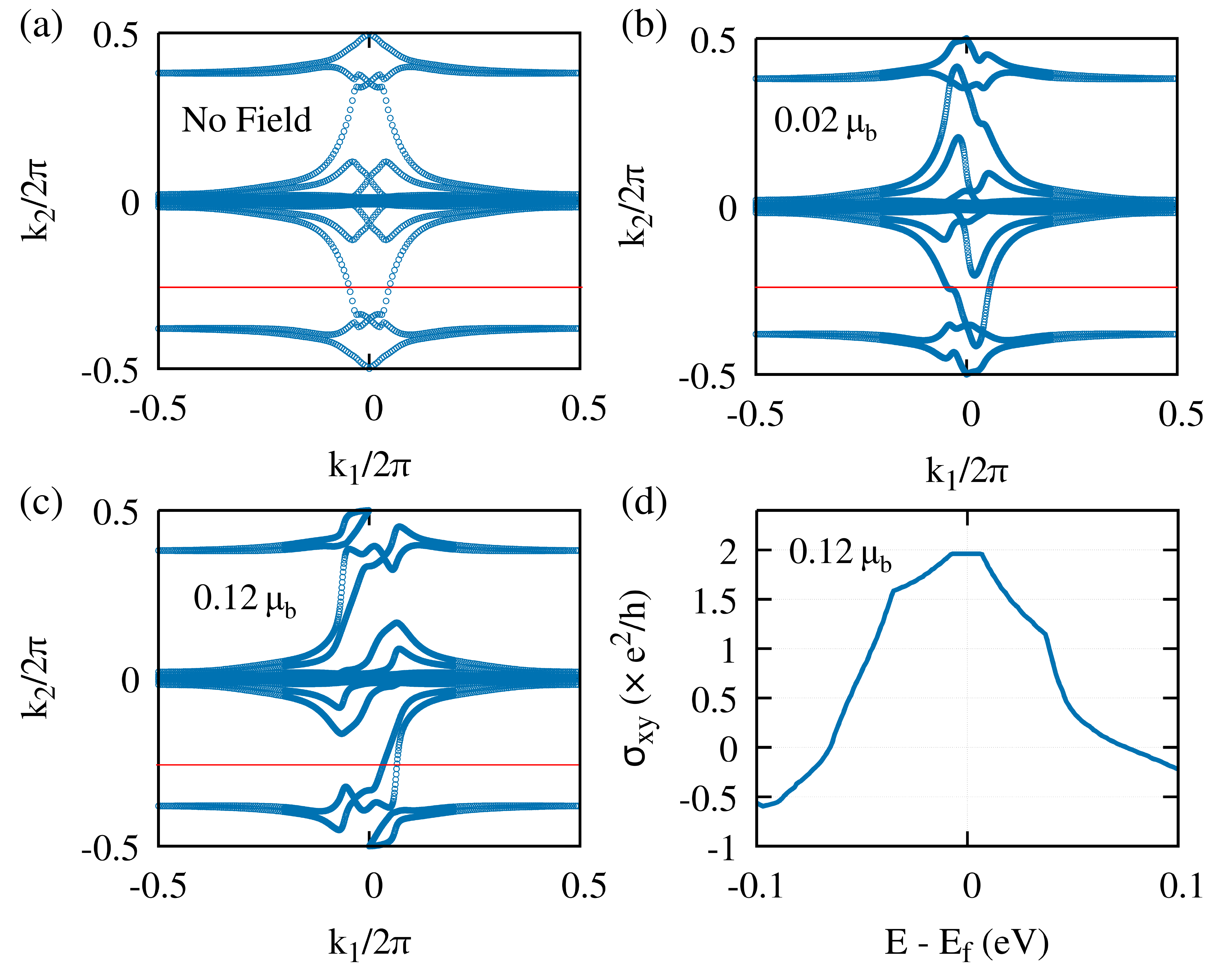}
		\caption{ Wannier center evolution in different conditions.			
			(a-c) Wannier center evolution under different magnetic fields.
			(d) Anomalous hall conduct of the QH state in the unit of $\text{e}^2$/h
			under different chemical potentials.}
		\label{fig4}
	\end{figure}

	The magnetic field serves as an effective way to tune the electronic 
	band structure and band order, as it breaks the time-reversal symmetry.
	We try to apply an external magnetic field perpendicular to the monolayer
	$\text{TaIrTe}_4$, see the sketch in Fig.~\ref{fig1}(a). As long as a non-zero
	magnetic field is introduced, the degeneracy of spin-up and spin-down is broken and
	an obvious Zeeman split could be observed in both valence and
	conduction bands, see Fig.~\ref{fig3}(b). Correspondingly, the time-reversal symmetry
	of Wanneir center evolution is broken compared to the non-field state and 
	the positive and negative parts along $k_1$ are not equal anymore, see Fig.~\ref{fig4}(a-b).
	
	As the magnetic field increases, the Zeeman split gets larger. In addition to the 
	band spin split, we can see that the spin-up and spin-down channels move in opposite 
	directions in energy space, as the comparison between Fig.~\ref{fig3}(b-d). 
	In this mechanism, the original band inversion from orbital1-spin-down and orbital2-spin-up
	channels remain unchanged, only with the magnitude of the inverted band gap increasing. 
	On the other hand, the spin-up channel from orbital1 moves down, and the spin-down channel from orbital2
	moves up, a new band inversion from orbital1-spin-up and orbital2-spin-down happens,
	see the schematic diagram in Fig.~\ref{fig3}(e).
	
	From band number indexed Berry phase calculations, we found that the original 
	band inversion due to crystal field and spin-orbital coupling, i.e. the band inversion between 
	orbital1-spin-down and orbital2-spin-up, hosts a Chern number 1. Similarly,
	the original band inversion between orbital1-spin-up and orbital2-spin-down
	hosts a Chern number -1. Hence, in the case of nontrivial $Z_2$ state
	and with a weak Zeeman field, the net Chern number is 0. On the other hand,
	the newly generated band inversion between orbital1-spin-up and orbital2-spin-down
	induced from the magnetic field has a Chern number 1. Consequently,  
	the new phase with  Zeeman field above 0.1$\mu_B$ has a none zero quantized 
	Hall conductance of 2$e^{2}/h$.
	
	The evolution of band gaps between different orbital and spin characters,
	i.e. $\Delta E_1=E_{or1-down}-E_{or2-up}$ and $\Delta E_2=E_{or1-up}-E_{or2-down}$, is given in 
	Fig.~\ref{fig3}(f). We can see that the positive band gap of $\Delta E_2$
	decreases to zero at around $gB_{z}=0.08~\mu_B$ and then goes to the negative zone. 
	Correspondingly, the Chern number between orbital1-spin-up and orbital2-spin-down
	changes from -1 to 1. On the other hand, the band gap of $\Delta E_1$
	follows the opposite trend. It keeps staying at the positive zone and shares the same trend
	with the increasing of magnetic field up to 0.12 $\mu_B$, with a constant Chern number 1
	in the whole process. Therefore, the topological phase transition is mainly induced by 
	the band order between orbital1-spin-up and orbital2-spin-down. The sign change of 
	Chern number from  -1 to 1 for the band gap of $\Delta E_2$ leads to a 
	net Chern number 2 for the whole system.
	
	The QH insulating state was also confirmed by the Wannier center evolution
	in the 2D BZ at the magnetic field of $gB_z=0.12~\mu_B$. As presented in
	Fig.~\ref{fig4}(c), with time-reversal symmetry breaking, the evolution of 
	Wannier center evolves in the whole range of $k_1$ axis. When fixing the band 
	number n as the fully occupied states, the Wannier center evolution lines 
	cross the reference line twice, where both of the two evolution lines show 
	a positive slope. From the energy dispersion in Fig.~\ref{fig3}(d), we can see 
	a global band gap around 14 meV, so a quantized Hall conductance 
	is expected. We then further calculated the chemical potential dependent 
	Hall conductance by following the linear response Kubo formula approach, 
	see the method part for details. From Fig.~\ref{fig4}(d),
	we can see a stable plateau of $2e^2/h$ near the charge neutrality point at the 
	situation of $gB_{z}=0.12~\mu_B$, fully in agreement with the band indexed Berry phase
	analysis and Wilson loop calculations. Based on the above analysis, we tried to 
	rotate the magnetic field to the $-z$ direction and found that both the slope
	of Wannier center evolutions and Hall conductance change the signs.
	
	\begin{figure}[htb]
		\centering
		\includegraphics[scale=0.15]{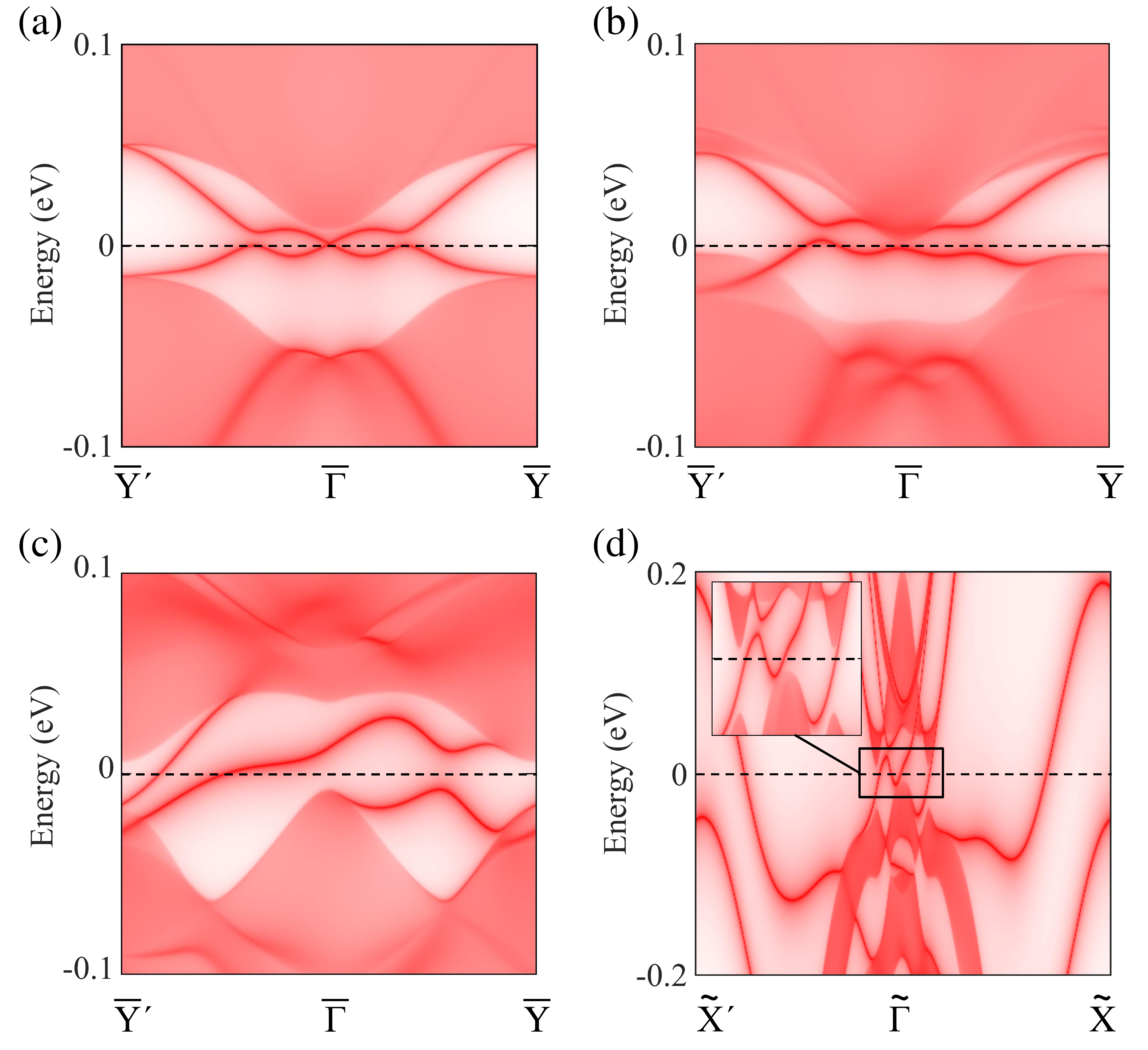}
		\caption{ Projected edge states in different situations for
			monolayer $\text{TaIrTe}_4$ with and without magnetic field.
			(a) [010] surface without magnetic field.
			(b) [010] surface with $gB_z=0.02~\mu_B$.
			(c) [010] surface with $gB_z=0.12~\mu_B$.
			(d) [100] surface with $gB_z=0.12~\mu_B$.
			The inset in (d) is an enlarged surface states near 
			$\tilde{\Gamma}$. The other edge states away from $\widetilde{\Gamma}$
			could be attributed to the topological order of intra-valance bands.}
		\label{fig5}
	\end{figure}%
	
	For both QSH insulators and QH insulators, the quantized signals originate from 
	the edge states located inside the bulk band gaps. Fig.~\ref{fig5}(a) is the 
	projected edge states for the QSH insulator phase along  
	$\overline{Y}-\overline{\Gamma}-\overline{Y}$, with spin helical linear crossing 
	edge Dirac point locating at $\overline{\Gamma}$. As long as a $z$-oriented magnetic
	field is applied, the Dirac point is broken by opening an anti-crossing-like band gap,
	due to time-reversal symmetry breaking, see Fig.~\ref{fig5}(b). After the new band
	inversion happens between orbital1-spin-up and orbital2-spin-down at $X$ point, the 
	spin helical edge states transfer to chiral edge states with positive velocity connecting
	the occupied and non-occupied bulk bands, see Fig.~\ref{fig5}(c-d). Though the specific 
	shapes of the edge states are dependent on the details of edge potentials, the net crossing points
	between the chiral edge states and Fermi levels are both 2 for the open boundary condition
	along $x$ and $y$ directions, fully consistent with the bulk topological charge analysis.
	Therefore, the quantizeCd signals are switchable between longitudinal conductance and 
	transversal Hall conductance, via tuning the external magnetic field.

	\section{Conclusion}
	In summary, we studied magnetic field-induced topological phase 
	transition in the newly discovered dual QSH insulator $\text{TaIrTe}_4$.
	Applying a magnetic field along $z$-direction, the original band order
	of the inverted band gap from one branch of the spin-up and spin-down channels
	is exchanged, with the corresponding Chern number transfering from -1 to 1.
	Together with another original branch of band inversion with Chern number
	1, the topological phase transition happens between the QSH insulator and the QH 
	insulator. Since the QSH insulator state in monolayer $\text{TaIrTe}_4$
	can host a non-zero quantized longitudianl conductance that rarely exists 
	in other 2D $Z_2$ topological insulators, such topological phase transition is 
	along with the exchange of quantized signals between longitudianl conductance 
	and transversal Hall conductance. This result also proposes another effective
	strategy to verify the existence of QSH insulating state in $\text{TaIrTe}_4$.

	\section{Method}
	Ground state study of monolayer $\text{TaIrTe}_4$ and 
	the Wannier projection is carried out
	in Full-Potential Local-Orbital(FPLO) package under generalized 
	gradient approximation (GGA) with Perdew-Burke-Ernzerhof (PBE) 
	parametrization~\cite{opahle1999,koepernik1999,koepernik2023}.
	Self-consistent energy reaches a convergence of $\text{10}^{-6}$ eV.
	Structure from the experimental is applied with lattice constants of
	a = 12.42~$\text{\AA}$ and b = 3.77~$\text{\AA}$, a vacuum of 15~$\text{\AA}$ is applied on
	c axis to eliminate the inter-layer interaction.
	Magnetic field is simulated by adding Zeeman splitting Hamiltonian
	on Wannier basis $H_0$ which reads $H = H_0 + H_Z$, where
	$H_Z=g\bm{B}\cdot\bm{\sigma}$. For out-of-plane magnetic
	field $\bm{B}\parallel$ z, it could be written as a scalar $H_Z=gB_z\sigma_z$.
	Based on $H$, edge states calculation is performed by the iteration of 
	green function\cite{mplopezsancho1984} while the wannier center evolution 
	is carried out by the Wilson loop method\cite{yu2011}.
	Anomalous hall conductivity with varying chemical potential is performed 
	by Kubo formula\cite{nagaosa2010} in clean limit
	\begin{equation}
		\begin{aligned}
			&\sigma_{xy}(E)=\frac{e^2}{\hbar S}\int[d\bm{k}]\sum_{i} f^i(E,\bm{k}) \Omega_{xy}^i(\bm{k})\\
			&\Omega_{xy}^i(\bm{k})=\sum_j Im\{r_x^{ij}(\bm{k}),r_y^{ji}(\bm{k})\} \\ 
		\end{aligned}
	\end{equation}
	whereas $r_a^{ij}(\bm{k})=i\braket{u_i(\bm{k})|\partial_{k_a}|u_j(\bm{k})}$, 
	$\ket{u_i(\bm{k})}$ represents the $\text{i}^{\text{th}}$ Wannier state, 
	S is the in-plane area of monolayer $\text{TaIrTe}_4$, 
	$H_{ij}(\bm{k})$ is the Hamiltonian of the system while
	$f^i(E,\bm{k})$ is the occupation of band index i with momenta $\bm{k}$ 
	under a chemical potential E.
	For the numerical integration over BZ, a $k$-point sampling of 6$\times$18$\times$1
	is applied for the ground state DFT study while a sampling of 2000$\times$2000$\times$1
	is used for the anomalous hall conductivity calculation.
	
	\begin{acknowledgement}
		
		This work was supported by the National Key R\&D Program of China
		(Grant No. 2021YFB3501503), the National Natural Science
		Foundation of China (Grants No. 52271016 and No. 52188101),
		and Foundation from Liaoning Province (Grant No. XLYC2203080). 
		Part of the numerical calculations in this study were carried out on
		the ORISE Supercomputer.
		
	\end{acknowledgement}
	\providecommand{\latin}[1]{#1}
	\makeatletter
	\providecommand{\doi}
	{\begingroup\let\do\@makeother\dospecials
		\catcode`\{=1 \catcode`\}=2 \doi@aux}
	\providecommand{\doi@aux}[1]{\endgroup\texttt{#1}}
	\makeatother
	\providecommand*\mcitethebibliography{\thebibliography}
	\csname @ifundefined\endcsname{endmcitethebibliography}
	{\let\endmcitethebibliography\endthebibliography}{}

	% \bibliography{qhtex.bib}% Produces the bibliography via BibTeX.
	
\end{document}